\documentclass[12pt]{article}
\usepackage[english]{babel}
\usepackage[dvipsnames,svgnames,x11names]{xcolor}
\usepackage[marginparwidth=2.5cm]{geometry}
 \usepackage[final]{changes}

\usepackage{todonotes}
\usepackage{graphicx}
\usepackage{makeidx}
\usepackage{multirow}
\usepackage{amssymb}
\usepackage{amsfonts}
\usepackage{amsmath}
\usepackage{amsthm}
\usepackage{natbib}

\usepackage{graphicx}% Include figure files
\usepackage{dcolumn}% Align table columns on decimal point
\usepackage{bm}% bold math
\usepackage[utf8]{inputenc}
\usepackage[T1]{fontenc}
\usepackage{mathptmx}
\usepackage{amsfonts}
\usepackage{bbm}
\usepackage[bottom]{footmisc}

\begin{document}

\title
{Quantum probability for statisticians; some new ideas\\ Running head: Quantum probability}

\author{Inge S. Helland\\ Department of Mathematics, University of Oslo\\P.O.Box 1053, N-0316 Oslo, Norway\\ ingeh@math.uio.no\\Orcid: 0000-0002-7136-873X}

\date{}

\maketitle

\begin{abstract}
It is argued from several points of view that quantum probabilities might play a role in statistical settings. New approaches toward quantum foundations have postulates that appear to be equally valid in macroscopic settings. One such approach is described here in detail, while one other is briefly sketched. In particular, arguments behind the Born rule, which gives the basis for quantum probabilities, are given. A list of ideas for possible statistical applications of quantum probabilities is provided and discussed. A particular area is machine learning, where there exists substantial literature on links to quantum probability. Here, an idea about model reduction is sketched and is motivated from a quantum probability model. Quantum models can play a role in model reduction, where the partial least squares regression model is a special case. It is shown that for certain experiments, a Bayesian prior given by a quantum probability can be motivated. 

\end{abstract}


\section{Introduction}

The basis for nearly all articles in theoretical and applied statistics is Kolmogorovian probability (Kolmogorov, 1933). Quantum probability is mostly looked upon by statisticians as an exotic tool with no relevance for statistical science and statistical inference. This implies that, in the statistical literature there is very little discussion of possible links between statistical theory and quantum theory. (A good exception is the article  Barndorff-Nielsen et al. (2003), where quantum versions of exponential models, sufficiency and Fisher information are discussed, among other related topics.)

In my opinion, this has led to a limitation of statistical science, both from a theoretical point of view and from an applied point of view. I relate this statement to the idea that traditional statistical theory only - at least in most cases - looks upon the parameter space as a set with no structure. Introducing more structure into the parameter space will give a richer theory. And I will show in the next Section that some such structures indeed lead to quantum probabilities. Later in this article I will discuss a specific structure, symmetry introduced by letting a concrete group act on the parameter space and the implications for model reduction.

It seems as if a discussion of links between quantum theory and statistical theory is more than due now. This can be underpinned by looking at the current revolution in artificial intelligence, in particular machine learning, an area that engages more and more statisticians. It is probably not so well known among statisticians that discussions of connections between machine learning and quantum mechanics have appeared in the literature now; see for instance the review article Dunjko \& Briegel (2019). If we really mean that machine learning should be based upon statistics, connections between statistical theory and quantum theory should also be of high interest.

One can ask then: Is not quantum probabilities just of relevance to the microcosmos? My answer to this is no. To argue for that answer, one can look at recent derivations of quantum mechanics from various postulates. In two books (Helland, 2021; Helland and Parthasarathy, 2024) and in a series of articles (Helland 2022a,b, 2023a-d, 2024a-f), this author has proposed a completely new foundation of quantum theory, and also, in this connection, discussed the interpretation of the theory. This foundation is summarized in Subsections 2.2 and 2.3 below, where the essence of quantum theory is deduced from 7 postulates. Looking at these postulates, most of them may be seen to be  equally valid in a macroscopic setting.

Another basis for quantum theory is given in a series of articles by De Raedt and his collaborators. One of these is De Raedt et al. (2014). Here, essential elements of quantum mechanics are derived from what is called logical inference to experiments, an assumption that there are uncertainties about individual events and that the frequencies are robust with respect to small changes in the parameters. One tool in the derivations is that of Fisher information. Fisher information as a general basis for nearly all aspects of science has earlier been advocated by Frieden  (2014).

Finally, it is of interest in this connection to look at the quantum-like models introduced by Khrennikov and collaborators; see Khrennikov (2010, 2021) and Haven \& Khrennikov (2013, 2016). These models are based upon quantum probabilities  and are applied to several sciences, including cognitive psychology, sociology, finance and biology. An important element here is a theory of quantum decisions; see also Helland (2023d, 2024e).

While there are few examples of statistical papers related to quantum probability, there are  several articles in the quantum theory literature that use basic statistical concepts and ideas. Very much discussion among theoretical physicists is concerned with the interpretation of quantum theory. Some still want to look upon it as an ontological theory, but more and more physicists now conclude with the view that quantum mechanics should be interpreted as an epistemic or epistemological theory: It is a theory of our knowledge of the world, in the same way that statistical theory is basically about knowledge from experiments and observations. A strong school in the quantum community is QBism, originally an abbreviation for Quantum Bayesianism, but now by its founders claimed to have a somewhat wider basis, see Caves et al. (2002) and Fuchs (2003, 2010). A very recent article on Bayes' rule and related inference in quantum mechanics is Liu (2024).

The plan of this article is as follows: In Section 2 my own basis for quantum theory is discussed in some detail, together another basis. One aim is to motivate the use of quantum probabilities in macroscopic settings. In Section 3, several potential ideas for the use of quantum probabilities in statistical settings are discussed. This includes model reductions, quantum probabilities as priors for certain experiments, and machine learning. The final section gives some general discussion points.

\section{Two basic theories leading to quantum probabilities}
\label{sec2}

\subsection{A new approach; the postulates and the first results}

A completely new approach towards quantum foundations is proposed in Helland (2024a-d), where the formal properties of quantum mechanics are derived from a rather simple set of postulates. These postulates will be repeated below. It should be emphasizes that the postulates imply essential parts of the ordinary formalism, in particular the formalism behind quantum probabilities. The purpose of the postulates is to imply an increased understanding of the quantum formalism.

As a possible general interpretation, the basis can be taken to be relative to an actor who is in some fixed (physical or statistical) context. In this context there are \emph{theoretical variables}, and some of these variables, say $\theta, \lambda, \eta,...$ may be related to the actor $C$. Some of these variables are \emph{accessible} to him, which means roughly that it is, in some future, given some estimation principle, in principle possible to obtain as accurate estimates as he wishes on the relevant variable. Other variables are \emph{inaccessible}. A typical physical example of an inaccessible variable is the vector (position, momentum) of a particle, a variable which by Heisenberg's inequality can not be given an accurate estimate. In Helland (2024a-d), other physical examples are given. In the present article, applications to statistics will be the theme, and then in most cases the theoretical variables will be parameters of some statistical model. However, I will also allow other interpretations of the variables: In special applications they may be latent variables, future data, or combinations of parameters and data. 

In the parameter case, an inaccessible parameter may be a total parameter in a very general model where this total parameter can not be estimated. Another application will be to machine learning, where the hidden variables may be looked upon as inaccessible.

The above characterization of accessible and inaccessible variables will in this article thus mainly be related to a  statistical  implication of the theory. But the theory \added{itself} is purely mathematical and can be made precise in different directions. In particular, the terms `accessible' and `inaccessible' can just be seen as primitive notions of the theory. In addition to the statistical implication, two other ways that the theory can be made precise are 1) ordinary quantum mechanics, where the theoretical variables are physical variables, in my interpretation connected to a fixed context and also to some actor; 2) quantum decision theory, where the variables are decision variables.  From a mathematical point of view, it is only assumed that if $\lambda$ is a theoretical variable and $\theta=f(\lambda)$ for some function $f$, then $\theta$ is a theoretical variable. And if $\lambda$ is accessible, then $\theta$ is accessible.

In physical applications the variables may also be connected to some actor in the sense that the actor observes the physical variables. Note, however, that actors may communicate. The mathematical model developed in the articles mentioned above is equally valid relative to a group of people that can communicate about the various theoretical variables. This gives a new interpretation of the theory, an interpretation where all theoretical variables are defined jointly for such a group of actors. In physical applications, the actor or the communicating group of actors is important. In statistical applications we may take the group to be the set of all possible actors; see Section 3 below.

From a mathematical point of view, an accessible variable $\theta$ is called maximal if there is no other accessible variable $\lambda$ such that $\theta = f(\lambda)$ for some non-invertible function $f$. In other words, the term `maximal' will then be seen to be maximal with respect to the partial ordering of variables given by $\alpha\le\beta$ iff $\alpha=f(\beta)$ for some function $f$.

To be precise, every accessible variable is assumed to vary over some topological space. In most cases they are real-valued or vector-valued, and all functions discussed are assumed to be Borel-measurable.

One possible assumption in my theory is that there exists an inaccessible variable $\phi$ such that all the accessible variables can be seen as functions of $\phi$. In simple physical applications, such a $\phi$ may easily be defined explicitly. In statistical applications, $\phi$ may be some total, inaccessible parameter, say, the set of all parameters that in some way may be included in a certain statistical model. 

As a summary of the above discussion, here are the first 3 postulates of the theory:
\bigskip

\textbf{Postulate 1:} \textit{If $\eta$ is a theoretical variable and $\gamma = f(\eta)$ for some function $f$, then $\gamma$ is also a theoretical variable.}

\bigskip

\textbf{Postulate 2:} \textit{If $\theta$ is accessible to $C$ and $\lambda= f (\theta)$ for some function $f$ , then $\lambda$ is also accessible to $C$.}
\bigskip

\textbf{Postulate 3:} \textit{In the given context there exists an inaccessible variable $\phi$ such that
all the accessible ones can be seen as functions of $\phi$. There is a group $K$ acting upon $\phi$.}
\bigskip

A definition is now needed for the fourth postulate:
\bigskip

\textbf{Definition 1 }
\textit{The accessible variable $\theta$ is called \emph{maximal} if $\theta$ is maximal as an accessible variable under the partial ordering defined by $\alpha\le \beta$ iff $\alpha=f(\beta )$ for some function $f$.}
\bigskip

Note that this partial ordering is consistent with accessibility: If $\beta$ is accessible and $\alpha=f(\beta )$, then $\alpha$ is accessible. Also, $\phi$ from Postulate 3 is an upper bound under this partial ordering. 
\bigskip

\textbf{Postulate 4:} \textit{There exist maximal accessible variables relative to this partial ordering. For every accessible variable $\theta$ there exists a maximal accessible variable $\lambda$ such that $\theta$ is a function of $\lambda$.}
\bigskip

Then, in my opinion, two different maximal accessible variables come very close to what Bohr called complementary variables; see Plotnitsky (2013) for a thorough discussion. The term complementary originated in connection to the variables position and momentum, but has now reached a number of applications; see Steiner \& Rendell (2024) and Maccone (2024) for examples.

It is crucial what is meant by `different' here. If $\theta=f(\eta)$, where $f$ is a bijective function, (i.e., there is a one-to-one correspondence between $\theta$ and $\eta$), then $\theta$ and $\eta$ contain the same information, and they must be considered `equal' in this sense. $\theta$ and $\eta$ are said to be `different' if they are not `equal' in this meaning. This is consistent with the partial ordering in Definition 1. The word `different' is used in the same meaning in the Theorem below. In other papers, the more precise term `non-equivalent' is used instead of `different' here.

Postulate 4 can be motivated by using Zorn’s lemma - if this lemma, which is equivalent to the axiom of choice, is assumed to hold - and Postulate 3, but such a motivation is not necessary if Postulate 4 is accepted. Physical examples of maximal accessible variables
are the position or the momentum of some particle, or the spin component in some
direction. In a more general situation, the maximal accessible variable may be a vector, whose components are simultaneously measurable variables.

Assuming these postulates, the main result of Helland (2022a, 2024a,d) is as follows:
\bigskip

\textbf{Theorem 1} \textit{Consider a context where there are two different maximal accessible variables $\theta$ and $\eta$. Assume that both $\theta$ and $\eta$ are real-valued or real vectors, taking at least two values.  Make the following additional assumptions:}
\smallskip

\textit{(i). On one of these two variables, say $\theta$, there can be defined a transitive group of actions $G$ with a trivial isotropy group and with a left-invariant measure $\rho$ on the space $\Omega_\theta$, the space upon which $\theta$ varies.}

\textit{(ii) There exists a group $K$ on $\Omega_\phi$.}

\textit{(iii) The ranges of the two variables are similar, meaning that there exists a binary function between the two ranges.}

\textit{Then there exists a Hilbert space $\mathcal{H}$ connected to the situation, and to every (real-valued or vector-valued) accessible variable there can be associated a symmetric operator on $\mathcal{H}$.}
\bigskip

In Helland (2026a), it is proved that Postulate 3 and Assumption (ii) are not needed in this Theorem.

 The main result here is that each accessible variable $\xi$ is associated with an operator $A^\xi$. The proof goes by first constructing $A^\theta$ and $A^\eta$, then operators associated with other accessible variables are found by using the spectral theorem. For this we need weak conditions (Hall, 2013; Helland, 2024d) ensuring that the symmetric operators are self-adjoint.
\bigskip

In order to formulate in general the spectral theorem, first, the \emph{spectrum} $\sigma^A$ of an operator $A$ is defined as the set of constants $\lambda$ such that $A-\lambda I$ does not have a bounded inverse. For self-adjoint operators, the spectrum is contained in the real line and contains all eigenvalues of $A$. For bounded operators, the spectrum is equal to the set of eigenvalues.

Then in general we have for any self-adjoint operator $A$ that there exists a projection-valued measure $E^A$ such that
\begin{equation}
A=\int_{\sigma^A}\lambda dE^A (\lambda).
\label{spectral1}
\end{equation}

From this spectral theorem (see Hall (2013) for a proof), if $A=A^\eta$ is the operator which according to Theorem 1 is associated with the maximal accessible variable $\eta$, we can define the operator associated with $\xi=f(\eta)$ by
\begin{equation}
A^\xi = f(A)=\int_{\sigma^A}f(\lambda) dE^A (\lambda).
\label{spectral2}
\end{equation}

In particular, we have
\begin{equation}
\int_{\sigma^A} dE^A (\lambda)=I.
\label{spectral3}
\end{equation}

Note that here $\eta$ may be any maximal accessible variable associated with some $\theta$ which satisfies (i) and (iii) above. 

By Postulate 4, for any accessible variable $\xi$, there exists a maximal variable $\eta$ and a function $f$ such that $\xi=f(\eta)$. In this way, operators associated with any accessible variable may be defined.

Groups as acting on a space are important in my approach. A group action $G$ acting on a space $\Omega$ is called transitive if for every $\omega\in\Omega$, the range of $g\omega$ as $g$ runs through $G$ is the full space $\Omega$. If this holds for one $\omega$, it holds for every $\omega\in\Omega$. The isotropy group of $G$ at $\omega$ is the set of $g\in G$ such that $g\omega = \omega$. In the transitive case, for different $\omega$, the isotropy groups are isomorphic. In particular, if the isotropy group is trivial for one $\omega$, it is trivial for every $\omega\in\Omega$.

When there is a transitive group $G$ with a trivial isotropy group acting on $\Omega$, there will be a one-to-one correspondence between the points $\omega\in \Omega$ and the group elements $g\in G$. This is important for the proof of Theorem 1.

\bigskip

An important special case of Theorem 1 is when the accessible variables take a finite number of values, say $u_1,u_2,...,u_r$.  For this case it is proved in Helland (2024a,d)  that a group $G$ and a transformation $k$ with the above properties can always be constructed. The following Corollary then follows: 
\bigskip

\textbf{Corollary 1} \textit{Assume that there exist two different maximal accessible variables $\theta$ and $\eta$, each taking $r$ values, and not in one-to-one correspondence. Then, there exists an $r$-dimensional Hilbert space $\mathcal{H}$ describing the situation, and every accessible variable in this situation will have an associated self-adjoint operator in $\mathcal{H}$.}
\bigskip

In the finite case, the equations (\ref{spectral1}-\ref{spectral3}) take a simpler form. The operator $A^\eta$ will have eigenvalues $\{u_j\}$ and corresponding eigenvectors $\{\bm{u}_j\}$. The spectral theorem then reads
\begin{equation}
A^\eta = \sum_j u_j \bm{u}_j \bm{u}_j^\dagger,
\label{spectral4}
\end{equation}
where $\bm{u}_j^\dagger$ is the complex conjugate row vector corresponding to $\bm{u}_j$. In the quantum mechanical literature, these vectors are often written as ket and bra vectors: $\bm{u}_j =|j\rangle$ and $\bm{u}_j^\dagger = \langle j|$. The eigenvectors can be chosen as orthonormal: $\bm{u}_i^\dagger \bm{u}_j = \langle i |j\rangle =\delta_{ij}$. In the following, both notations will be used.

We then further have
\begin{equation}
A^{f(\eta )}= \sum_jf(u_j )\bm{u}_j \bm{u}_j^\dagger,
\label{spectral5}
\end{equation}
in particular
\begin{equation}
\sum_j \bm{u}_j\bm{u}_j^\dagger =I,
\label{spectral6}
\end{equation}
the corresponding resolution of the identity.

Theorem 1 and its Corollary constitute the first steps in a new proposed foundation of quantum theory. In statistical applications, the variables (parameters) involved will in most cases have a continuous variation. However, continuous parameters can be approximated by a sequence of parameters taking a finite number of values; see subsection 5.3 in Helland (2021). In this way we may avoid both the symmetry assumptions of Theorem 1 and the technical issues relating symmetric and self-adjoint operators. Examples are given in Section 3 below.

The second step now is to prove the following: If $k$ is the transformation connecting two related maximal accessible variables $\theta$ and $\eta$, and $A^\theta$ and $A^\eta$ are the associated operators, then there is a unitary operator  $W(k)$ such that $A^\eta = W(k)^{-1} A^\theta W(k)$. This, and a more general related result is proved as Theorem 5 in Helland (2024a).

Given these results, a rich theory follows. The set of eigenvalues of the operator $A^\theta$ is proved to be identical to the set of possible values of $\theta$. The variable $\theta$ is shown to be maximal if and only if all eigenvalues of the corresponding operator are simple. In general, the eigenspaces of $A^\theta$ are in one-to-one correspondence with questions `What is $\theta$'/ `What will $\theta$ be if we measure it?' together with sharp answers $\theta=u$ for some eigenvalue $u$ of $A^\theta$. If $\theta$ is a maximal accessible variable, then all eigenvectors of the operator $A^\theta$ have a similar interpretation. In my theory, I wish to limit the concept of a state vector to vectors in the Hilbert space that are eigenvectors of some meaningful operator. Then this gives a simple interpretation of all possible state vectors.

In traditional quantum mechanics, a pure state is assumed to be given by a normalized vector $\mathbf{v}$ in the Hilbert space. In my theory, a pure state can be specified by a simple statement $\theta = u$, where $\theta$ is some maximal accessible variable. It follows from the discussion above:
\bigskip

\textbf{Qbservation}

\textit{There is an isomorphism betwee the statements of the form $\theta = u_i$ and the normalized eigenvectors $\mathbf{v}_i$ of the corresponding operators $A^\theta$.}
\bigskip

From this, it is natural to limit the state concept to ket vectors that are eigenvectors of some meaningful operator, a possible limitation of the superposition principle.

What so far has ben lacking in the above theory, is a foundation of Born's formula, necessary for the computation of quantum probabilities. In the literature, several authors have given conditions under which the Born formula can be derived. In particular, versions of Born's formula are proved under a weak condition in Helland (2026b). This condition is there seen as a consequence of
\bigskip

\textbf{Postulate 5:} \textit{The likelihood principle holds.}
\bigskip

As is well known, the likelihood principle is a principle that many statisticians base their inference on. In its strict form it is controversial, see, for instance, the discussion in Schweder and Hjort (2016). Elsewhere (Helland, 2021), I have advocated the view that the principle should be restricted to a specific context, and then it is less controversial. For a basic historical discussion of the principle, see Berger and Wolpert (1988). 

Recall that the likelihood principle runs as follows: \emph{Relative to any experiment, the experimental evidence is always a function of the likelihood.} Here, the term `experimental evidence' is left undefined and can be made precise in several directions. But as everybody would agree, an experiment is always done in a context, and such a context should include a well-defined experimental question.

In a quantum mechanical setting, a potential or actual experiment is seen in relation to an actor $C$ or to a communicating group of actors. Concentrate here on the first scenario. In the simplest case, assuming a discrete-valued variable, we assume that $C$ knows the state $|a;i\rangle$ of a physical system and that this state can be interpreted as the knowledge that $\theta^a = u_i$ for some maximal accessible variable $\theta^a$. Then assume that $C$ has focused upon a new maximal accessible variable $\theta^b$, and we are interested in the probability distribution of this variable.

In Helland (2021, 2024b, 2026b), a generalized likelihood principle is proved from the ordinary likelihood principle: Given some experiment, assumed here to have a discrete, maximal accessible parameter $\theta^a$, and assume a context $\tau$ connected to the experiment, any experimental evidence will under the above assumptions be a function of the so-called likelihood effect $F=F^a$, defined by 
\begin{equation}
F^a(\bm{u}; z,\tau)=\sum_i p(z|\tau, \theta^a=u_i)|a;i\rangle\langle a;i|,
\label{effect0}
\end{equation}.
where $ p(z|\tau, \theta^a=u_i )$ is the probability density or the point probability of the data.

In many textbooks, quantum mechanics is restricted to discrete-valued variables as above. For a continuous variable $\theta^a$, the likelihood effect can be defined by appealing to the spectral theorem for the operator $A=A^{\theta^a}$ and using the probability density $p(z|\tau,\theta^a=u)$ for the data:
\begin{equation}
F^a(\bm{u}; z,\tau)=\int_{\sigma^A} p(z|\tau, \theta^a=u) dE^A(u).
\label{effect1}
\end{equation}
\bigskip

\textbf{Proposition 1; the generalized likelihood principle}

\textit{ Assume the likelihood principle, and that $p(z|\tau, \theta^b=u_j)$ is a bijective function of $u_j$, and  $p(z|\tau, \theta^c=v_k)$ is a bijective function of $v_k$. Consider two potential experiments $b$ and $c$ in some context $\tau$,and assume that these have equal likelihood effects for two observations $z_1^b$ and $z_2^c$ in these two experiments. Then}

\textit{a) The relevant events in the two experiments are equal in the sense that one can use the same Hilbert space to describe the results of the experiments, and the  events $\theta^b = u_j$ and $\theta^c = v_k$ can be ordered such that projections in this Hilbert space of these events are equal.}

\textit{b) The observations  produce equal experimental evidence on the associated accessible variables $\theta^b$ and $\theta^c$.}

\bigskip

\underline{Proof}

This follows from the ordinary likelihood principle and Theorem 3 in Helland (2026b).

$\qed$

\subsection{The Born rule from the approach above}

Using these postulates and a version of Gleason's Theorem due to Busch (2003), the following variant of Born's formula is proved in Helland (2026b):
\bigskip

\textbf{Theorem 2} [Born's formula, simple version] \textit{ Assume Postulates 5 and 6, and assume two different discrete maximal accessible variables $\theta^a$ and $\theta^b$. In the above situation we have:}
\begin{equation}
P(\theta^b =v_{j} |\theta^a =u_{i})=|\langle a;i|b;j\rangle|^2 .
\label{Born}
\end{equation}

Here, \added{$|a;i\rangle$ is the state given by $\theta^a=u_i$ and} $|b;j\rangle$ is the state given by $\theta^b=v_j$. In this version of the Born formula, I have assumed \emph{perfect measurements:} there is no experimental noise, so that the experiment gives a direct value of the relevant theoretical variable. The cross-product $\langle a;i|b;j\rangle$ is often called a probability amplitude.
\bigskip

A last postulate is needed to compute probabilities of independent events. A version
of such a postulate is
\bigskip

\textbf{Postulate 6:} \textit{ If the probability of an event $E_1$ is computed by a probability amplitude $z_1$ from the Born rule in the Hilbert space $\mathcal{H}_1$, the probability of an event $E_2$ is computed by a probability amplitude $z_2$ from the Born rule in the Hilbert space $\mathcal{H}_2$, and these two events are
independent, then the probability of the event $E_1 \cap E_2$ can be computed from the probability
amplitude $z_1 z_2$, associated with the tensor product of the Hilbert spaces $\mathcal{H}_1$ and $\mathcal{H}_2$.}
\bigskip

This postulate can be motivated by its relation to classical probability theory: If $P(E_1) =
|z_1|^2 $ and $P(E_2) = |z_2|^2$, then $P(E_1 \cap E_2) = P(E_1)P(E_2) = |z_1|^2 |z_2|^2 = |z_1z_2]^2$
\bigskip

The simple Born formula can now be generalized to the case where the variables are continuous, and where the accessible variables  $\theta^a$ and $\theta^b$ are not necessarily maximal. There is also a variant for a mixed state involving $\theta^a$. (See also Helland, 2026b.)

Define first the mixed state operator associated with any accessible variable $\theta$ by using the spectral theorem for $A=A^\theta$:
\begin{equation}
\rho^\theta = \int_{\sigma^A} p^\theta (u) dE^A(u).
\label{rho1}
\end{equation}

In the continuous case, $p^\theta (u)$ is the probability density for $\theta$. In the discrete case, $p^\theta (u)$ is the point probability of $\theta$, and (\ref{rho1}) reads
\begin{equation}
\rho^\theta = \sum_j p^\theta (u_j) \bm{u}_j \bm{u}_j^\dagger,
\label{rho2}
\end{equation}
where $\bm{u}_j$ is the eigenvector of $A^\theta$ corresponding to the eigenvalue $u_j$.

The probability distribution for $\theta$ assumed in (\ref{rho1}) can be of many kinds. For a Bayesian, it can be a prior or posterior distribution. The prior distribution case may be relevant in any case; the posterior distibution case is relevant when one has a series of experiments. For a frequentist, the `probability distribution' can be a confidence distribution of the kind discussed by Schweder and Hjort (2016). Also, statisticians that follow some fiducial school operate with probability distributions of parameters.

Assume now first that $\theta^b$ in Theorem 2 is discrete, but not necessarily maximal. Then $\theta^b$ is a function $f$ of a maximal accessible variable $\eta$, and it follows by summation over $j$, assuming that $\eta=v_j$ belongs to some set $B_1$ defined by $f(v_j)\in B$ for some given set $B$, that
\begin{equation}
P(\theta^b \in B|\theta^a =i)=\langle a; i| \Pi_{B}| a:i \rangle,
\label{Born1}
\end{equation}
where $\Pi_B$ is the projection upon the space spanned by the eigenvectors $|b;j\rangle$ of $A^\eta$ for which the eigenvalues $v_j$ are in the set of indices $j$ such that $f(v_j)\in B$.

Now, by approximating a continuous $\theta^b$ by discrete variables $\theta^{bn}$ such that $\theta^{bn}\rightarrow\theta^b$ as $n\rightarrow\infty$, it is easy to show that (\ref{Born1}) holds in general, where now $\Pi_B$ is interpreted as the projection upon the eigenspace of the indicator for the set $\theta^b\in B$ corresponding to the value 1 for this indicator. More precisely, we should use $\Pi_B=\int_{B\cap \sigma^b}dE^b(u)$, where $\sigma^b$ is the spectrum of the operator $A^{\theta^b}$, and $\{E^b\}$ is found from the spectral theorem of the same operator.

In the same way, a continuous $\theta^a$ may be approximated by discrete $\theta^{ar}$, assumed to be functions $f_r$ of some maximal accessible variables $\xi_r$, replacing the variable $\theta^a$ in Theorem 2. Then, using the definition (\ref{rho1}) (which can be generalized to the case of non-maximal accessible variables), we can prove
\bigskip

\textbf{Theorem 3} [Born's formula, general version] \textit{Assume Postulate 5 and Postulate 6, and assume that we have two different accessible variables $\theta^a$ and $\theta^b$ in the relevant context. Assume that the knowledge of $\theta^a$ is given by the density matrix $\rho^a$. Then for any Borel set $B$ in $\Omega_{\theta^b}$ we have}
\begin{equation}
P(\theta^b \in B|\rho^a) = \mathrm{trace} (\rho^a \Pi_B).
\label{Born4}
\end{equation}
\bigskip

This result is not necessarily associated with a microscopic situation, a fact that I will come back to in examples in Section 3. 

As a corollary, we have
\begin{equation}
E(\theta^b|\rho^a)=\mathrm{trace}(\rho^a A^{\theta^b}).
\label{Born7}
\end{equation}

Here $A^{\theta^b}$ is the operator corresponding to $\theta^b$.
\bigskip

Finally, one can generalize to the case where the final measurement is not necessarily perfect. Let us assume future data $z^b$ instead of a perfect theoretical variable $\theta^b$, which is now taken to be the parameter of the experiment. Note that we only need the likelihood principle (together with postulate 6) for perfect experiments in order to prove that (\ref{Born4}) is valid. Then we can define an operator corresponding to $z^b$ by
\begin{equation}
A^{z^b}=\int_{\sigma^A} z^b p(z^b |\theta^b =u)dE^b (u),
\label{Born5}
\end{equation}
where $\{E^b\}$ is found from the spectral theorem used on the operator $A=A^{\theta^b}$. 
Then from the version (\ref{Born4}) of the Born formula, we obtain
\begin{equation}
E(z^b|\rho^a)=\mathrm{trace}(\rho^a A^{z^b}).
\label{Born6}
\end{equation}
and
\begin{equation}
P(z^b\in F) =\mathrm{trace}(\rho^a\int_{z\in (F\cap\sigma^{A^1})} dE^{z^b} (z)),
\label{Born9}
\end{equation}
where $\{E^{zb}\}$ is found from the spectral theorem used on the operator $A^1=A^{z^b}$.
\bigskip

Again, elementary quantum mechanics uses discrete data and discrete parameters, a setting unfamiliar to statisticians, but useful as an approximation. Then $p(z^b|\theta^b =u_i)$ is the point probability of the data, we define
\begin{equation}
A^{z^b}=\sum_i  p(z^b |\theta^b =u_i) \bm{u}_i\bm{u}_i^\dagger,
\label{Born10}
\end{equation}
and Born's formula gives $E(z^b|\rho^a)=\mathrm{trace}(\rho^a A^{z^b})$ and
\begin{equation}
P(z^b \in F) =\mathrm{trace}(\rho^a \sum_{z_j\in F} \bm{v}_j \bm{v}_j^\dagger ),
\label{Born11}
\end{equation}
where $\bm{v}_j$ is the eigenvector of $A^{z^b}$ corresponding to the eigenvalue $z_j$. One can usually assume that $A^{z^b}$ has distinct eigenvalues.

All this not only points to a new foundation of quantum theory, but it also suggests a general epistemic interpretation of the theory: Quantum Theory is not directly a theory about the world, but a theory about an actor's knowledge of the world. In particular, the probabilities in the Born formula can be interpreted as probabilities attached to a single observer, or to a communicating group of observers.

\subsection{Quantum theory as the most robust description of reproducible experiments}

In De Raedt et al. (2014, 2016, 2023) another approach to the foundation of quantum theory is discussed. This approach will be described very briefly here. 

First, the authors define inference-probability as any conditional probability satisfying the three rules

1) $P(A|Z)+P(\bar{A}|Z)=1$, where $\bar{A}$ denotes not-$A$.

2) $P(AB|Z)=P(A|BZ)P(B|Z)=P(B|AZ)P(A|Z)$, where $AB$ denotes essentially what statisticians call $A\cap B$.

3) $P(A\bar{A}|Z)=0$ and $P(A+\bar{A}|Z)=1$, where $A+B$ denotes essentially what statisticians call $A\cup B$.

These rules are the same as the rules for the concept of plausibility, derived from reasonable assumptions and discussed in detail by Jaynes (2003). In op. cit., $A$, $B$ and $Z$ are propositions, and to be precise, $AB$ denotes that both propositions $A$ and $B$ are true, and $A+B$ denotes that at least one of the propositions is true.

Next, De Raedt et al. (2014, 2016) assume the following conditions, which are made precise in these articles:
\bigskip

\textbf{Conditions 1}
 \textit{There may be uncertainty about each event. The condition under which the experiment is carried out may be uncertain. The frequencies with which events are observed are reproducible and robust against small changes in the conditions. Individual events are independent.}
 \bigskip

Using these assumptions, they first derive quantum probabilities for the Einstein-Podolsky-Rosen-Bohm thought experiment. This experiment consists of a source $S$, a router $R_1$ to the left of the source oriented by a chosen unit vector $\mathbf{a}_1$, a router $R_2$ to the right of the source oriented by a chosen unit vector $\mathbf{a}_2$, and two detectors connected to each router.

The experiment is the same as what is called the Bell experiment, an experiment that has now been shown experimentally to give outcomes as predicted by quantum theory, but not by `common sense' use of classical arguments. The Bell experiment has been discussed under various assumptions by many authors, including Helland (2022b, 2023d)

De Raedt aims at making a minimal set of assumptions about the experiment; for details, see De Raedt et al. (2014, 2016) and references therein:

a) Each time the source $S$ is activated, it sends a signal to the right router and a signal to the left router.

b) After this, each router sends a signal to one of its two detectors, depending upon the orientations $a_i$.

c) The detectors register the signal and operate with 100 $\%$ efficiency.

The whole experiment is then run a large number $N$ of times, giving frequencies for the various outcomes.  Probabilities are obtained by letting $N\rightarrow\infty$. It turns out that these probabilities are in agreement with quantum theory, which deviates from common sense use of classical arguments.

This is a special experiment, but it is an important experiment, distinguishing between quantum predictions and classical conditions. A simpler experiment is the Stern-Gerlach experiment. Here the source $S$, activated at times $n=1,2,...,N$, sends a particle carrying a magnetic moment $\mathbf{s}$ to a magnet $M$ with its magnetization $\mathbf{a}$. Depending on $\mathbf{a}$ and $\mathbf{s}$, the particle is detected with 100 $\%$ by one of two assumed detectors. It is crucial that this depends only on the scalar product $\mathbf{a}\cdot\mathbf{s}$. Again, by a long argument, the predictions of quantum theory are derived under very weak assumptions.

In addition to this, there is a discussion of the Schr\"{o}dinger equation in De Raedt et al. (2014), a theme that I will not discuss in detail in the present article. It is of some relevance, however, that the concept of Fisher information was used in this derivation.

It is crucial that fundamental quantum predictions are derived in op. cit., using plausible assumptions together with the basic Conditions 1. There is nothing microscopic connected to these assumptions, supporting my main thesis that quantum theory also may be relevant in a macroscopic context. 

Finally, these derivations are also consistent with my epistemic interpretation of quantum theory, as shown by the following citation from DeRaedt et al. (2014): `... current scientific knowledge derives, through cognitive processes in the human brain, from the discrete events which are observed in laboratory experiments and from the relationships between those experiments that we, humans, discover.'

In a recent, long and detailed article (De Raedt et al., 2023), the Bell experient and the violation of the so-called CHSH inequality is discussed from the general point of view of mathematical models and discrete data. It is concluded that discrete data recorded by experiments and mathematical models used to describe relevant features belong to different, separate universes and should be treated accordingly. This is a conclusion that seemingly also has consequences for ordinary statistical modelling. We are used to looking upon the sample space and the parameter space as separate spaces. My views on the CHSH inequality are given in Helland (2024b,d); see a brief discussion in Subsection 3.5 below.

\section{Statistical applications}

In the statistical applications below, I will only in some special cases go into details concerning the related quantum-mechanical calculations, which may be complicated. The main purpose of this Section is to point at some ideas under which such calculations may possibly enlighten or complement a statistical analysis.

\subsection{A classical experiment that can be analyzed by quantum probabilities}

In a medical experiment, let $\mu_{a}, \mu_{b}, \mu_{c}$ and $\mu_{d}$ be continuous inaccessible parameters, the hypothetical effects of treatment $a, b, c$ and $d$, 
respectively. Assume that the focus of the experiment is to compare treatment $b$ with the mean effect of the other treatments, which is supposed to give the parameter 
$\frac{1}{3}(\mu_{a}+\mu_{c}+\mu_{d})$. One wants to do a pairwise experiment, but it turns out that the maximal parameter which can be estimated is
\[\theta^b =\mathrm{sign}(\mu^b -\frac{1}{3}(\mu_{a}+\mu_{c}+\mu_{d})).\]
(Imagine for example that one has four different ointments against rash. A patient is treated with ointment $b$ on one side of his back; a mixture of the other 
ointments on the other side of his back. It is only possible to observe which side improves best, but this observation is assumed to be very accurate. One can in 
principle do the experiment on several patients and select out the patients where the difference is clear.)

Described in this way, it may be natural, after the data are collected, to do a Bayesian analysis with a prior given by $P(\theta^b =-1)=P(\theta^b = +1)=1/2$. But assume now that we have the following modification of the experiment:

The experiment is done on a selected set of experimental units, on whom it is known from earlier accurate experiments that the corresponding parameter
\[\theta^a =\mathrm{sign}(\mu^a -\frac{1}{3}(\mu_{b}+\mu_{c}+\mu_{d}))\]
takes the value $+1$. In other words, for a Bayesian analysis one is interested in the priors
\[\pi =P(\theta^b=+1|\theta^a=+1).\]
\[1-\pi =P(\theta^b=-1|\theta^a=+1).\]

Consider first a full Bayesian approach, also toward these priors. Natural priors for $\mu_{a},...,\mu_{d}$ are independent $N(\nu, \sigma^2)$ with the same $\nu$ and $\sigma$. By location 
and scale invariance, there is no loss in generality by assuming $\nu =0$ and $\sigma=1$. Then the joint prior of 
$\zeta^a =\mu_{a}-\frac{1}{3}(\mu_{b}+\mu_{c}+\mu_{d})$ and $\zeta^b =\mu_{b}-\frac{1}{3}(\mu_{a}+\mu_{c}+\mu_{d})$ is multinormal with mean 
$\bm{0}$ and covariance matrix
\[\left( \begin{array}{cc}\frac{4}{3}&-\frac{4}{9}\\
-\frac{4}{9}&\frac{4}{3}\end{array}\right).\]
A numerical calculation from this gives
\[\pi=P(\zeta^b>0|\zeta^a>0)\approx 0.43.\]
This result can also be assumed to be valid when $\sigma\rightarrow\infty$, a case which  can be considered as independent objective priors for 
$\mu_{a},...,\mu_{d}$, more precisely, a joint non-informative prior for the parameters under the translation group; see Helland (2004).

Now consider quantum probabilities for the same priors. We start by looking at symmetries of the situation. Since again scale is irrelevant, a natural group on $\mu_{a},...,\mu_{d}$ is a 4-dimensional rotation group 
around a point $(\nu,...,\nu)$ together with a translation of $\nu$. Furthermore, $\zeta^a$ and $\zeta^b$ are contrasts, that is, linear combinations with coefficients 
adding to 0. The space of such contrasts is a 3-dimensional subspace of the original 4-dimensional space, and by a single orthogonal transformation, the relevant subset 
of the 4-dimensional rotations can be transformed into the group $G$ of 3-dimensional rotations on this latter space, and the translation in $\nu$ is irrelevant. One such 
orthogonal transformation is given by
\[\psi_{0}=\frac{1}{2}(\mu_{a}+\mu_{b}+\mu_{c}+\mu_{d}),\]
\[\psi_{1}=\frac{1}{2}(-\mu_{a}-\mu_{b}+\mu_{c}+\mu_{d}),\]
\[\psi_{2}=\frac{1}{2}(-\mu_{a}+\mu_{b}-\mu_{c}+\mu_{d}),\]
\[\psi_{3}=\frac{1}{2}(-\mu_{a}+\mu_{b}+\mu_{c}-\mu_{d}).\]
Let $G$ be the group of rotations orthogonal to $\psi_{0}$. We find
\[\zeta^a=-\frac{2}{3}(\psi_{1}+\psi_{2}+\psi_{3}),\]
\[\zeta^b=-\frac{2}{3}(\psi_{1}-\psi_{2}-\psi_{3}).\]
The rotation group element transforming $\zeta^a$ into $\zeta^b$  under $G$ is strongly related to the group element $g_{ab}$ transforming 
$a=-\frac{1}{\sqrt{3}}(1,1,1)$ into $b=-\frac{1}{\sqrt{3}}(1,-1,-1)$ under a group of rotations of unit vectors.

Furthermore, let $G^a$ be the maximal subgroup of $G$ under which $\zeta^a$ is 
permissible. The following definition was given in Helland (2004, 2006) and is further discussed in these two articles:
\bigskip

\textbf{Definition 2} \textit{Let $G$ be a group acting upon a parameter $\eta$, and let $\zeta$ be a function of $\eta$. We say that $\zeta$ is permissible if $\zeta(\eta_1)=\zeta(\eta_2)$ implies $\zeta(g\eta_1)=\zeta(g\eta_2)$ for all $g\in G$.}
\bigskip

In general, if $\zeta$ is a permissible parameter in this way, one can define group actions $h\in H$ on $\zeta$ by $h(\zeta(\eta))=\zeta(g\eta)$ for $g\in G$.
\bigskip

The subgroup $G^a$ is here isomorphic with the unit vector transformation group of rotations around $a$ together with a reflection in the plane perpendicular to $a$. The action by the group $H^a$ induced on $\zeta^a$ by $G^a$ is just a reflection together with the unit element. 

Again, all these groups have their analogues in relation to the rotation group of unit vectors. 

In conclusion, the whole situation is completely equivalent to the spin-example discussed in many books, for instance Helland (2021), and may be assumed to satisfy the postulates of Subsection 2.2 above. This implies by an application of the Born rule (see Proposition 7 in Helland (2021)):
\[\pi=P(\mathrm{sign}(\zeta^b)=+1|\mathrm{sign(\zeta^a)=+1})=\frac{1}{2}(1+a\cdot b)=\frac{1}{3}.\]
\smallskip

So the two analyses give different results for the desired prior. Which solution should one recommend? Here is my opinion: Both solutions are based upon symmetries implied by group actions. The full Bayesian solution is based upon a prior distribution on the inaccessible parameters $\mu_a , \mu_b , \mu_c$ and $\mu_d$, which could be related to group actions upon these parameters. In the quantum solution, one ends up with a group acting upon the accessible parameters $\theta^a = \mathrm{sign}(\zeta^a)$ and $\theta^b = \mathrm{sign}(\zeta^b)$, and priors based upon $\theta^a$ and $\theta^b$. In general, in applied statistics it is crucial that the parameter space is not too large. From an applied point of view, prior probabilities based upon accessible parameters should be preferred when compared to rather abstract probabilitiess based upon inaccessible parameters. Therefore, even though the arguments are more complicated, I will here prefer the quantum probability solution. Related arguments are discussed more generally in the next Subsections.

\subsection{Model reduction under symmetry}

In applied statistics, it is crucial that the parameter space is not too large. For instance, in regression problems, when the number of regression variables $p$ is larger than the number of units $n$, ordinary least squares regression runs into problems. Let in general $\phi$ be the set of all thinkable parameters that a statistician $C$ wants to include in his model, and, for the sake of the argument, let us assume that there exists a group $K$ acting on the space $\Omega_\phi$.

The total parameter $\phi$ may in many cases be so large that it cannot be estimated from the available data, using some estimation principle like unbiased estimation or equivariant estimation. Here, I will concentrate on the last estimation principle.

In general, let a group $G$ act upon the space $\Omega_\theta$ over which a parameter $\theta$ varies. In many cases, this group may be induced by a group $\widehat{G}$ acting upon the sample space $\Omega$, based upon some statistical model $P^\theta (x)$:
\begin{equation}
P^{g\theta}(\widehat{g}x)=P^\theta (x)\ \mathrm{for\ all}\ x\in \Omega.
\label{modelg}
\end{equation}

This introduces a \emph{homomorphism} from $\widehat{G}$ to $G$: If $\widehat{g}_1$ is mapped to $g_1$ and $\widehat{g}_2$ is mapped to $g_2$, then $\widehat{g}_1\widehat{g}_2$ is mapped to $g_1g_2$.

An estimator $\widehat{\theta}(x)$ of the parameter $\theta$ is said to be \emph{equivariant} if $\widehat{\theta}(\widehat{g}x)$ is the estimator of $g\theta$ whenever $\widehat{g}$ is mapped to $g$. There can be many arguments given to concentrate upon equivariant estimators.

Under very weak conditions there exists a left-invariant measure $\mu$ on $\Omega_\theta$ under the group $G$:  . The measure $\mu$ is said to be left-invariant if $\mu(A)=\mu (gA )$ for all $g$ and $A$. Right-invariant measures have a similar definition. In many cases, the left-invariant measure is equal to the right-invariant measure,

This introduces an invariant measure on every \emph{orbit} of the group $G$: An orbit is the set $\{g\theta_0\}$ for some $\theta_0 \in \Omega_\theta$. The space $\Omega_\theta$ is always divided into a disjoint set of orbits. If $\theta_1$ and $\theta_2$ belong to the same orbit, this orbit can equivalently be characterized by either $\{g\theta_1\}$ or $\{g\theta_2\}$. If there is only one orbit of $G$ in $\Omega_\theta$, the group is said to be \emph{transitive}.

An objective Bayes estimator with respect to $G$ is an estimator that uses the left-invariant measure or the right-invariant measure as a prior. In Helland (2004), 12 different reasons for using such a prior are given, among other things, it can be proved that credibility sets with some credibility probability are equal to frequentist  confidence sets with the same confidence probability.

Turning to quantum theory, it is important for the foundation that there exists a transitive group on the variable space  (see point (i) in Theorem 1 of Subsection 2.1). If $G$ should not be transitive, we can introduce the following model reduction principle:
\bigskip

\textbf{Principle 1} \textit{Reduce $\Omega_\theta$ to an orbit of the group $G$. Choose the orbit such that a subparameter $\zeta$ of interest is permissible.}
\bigskip

The property of being permissible was defined in Definition 2 of Subsection 3.1. The mapping from $g\in G$ to $h\in H$ given by $h\zeta(\theta)=\zeta(g\theta)$ is a homomorphism.

In Helland (2008) and Helland (2021), this principle is used on the electron spin, a qubit. It is shown that a classical model of spin can be reduced to a quantum model using this principle. In Helland (2001), the same principle was used to motivate the model reduction in multiple regression leading to the partial least squares regression model.

Go back to the example in Subsection 3.1. By a change of notation, let $G$ be the group given by reflections of three-dimensional unit vectors $\mathbf{a}$ together with rotations around  $\mathbf{a}$. This group is intransitive, and its orbits are found by fixing some  $\mathbf{a}$. This corresponds to what was called $G^a$ there, and a group reduction gives the quantum-mechanical interpretation of the example.

More statistical theory related to transitive and intransitive groups defined on the parameter space and the sample space are given in Helland (2004). It is of independent interest that the statistical model corresponding to partial least squares regression can be motivated by model reduction to orbits of a certain group defined on the parameter space (Helland, 2001, Helland, 2024f) .

\subsection{Partial least squares regression and model reduction}

Partial least squares regression is an algorithmic method for estimating the regression parameter $\beta$, intended for the case of collinearity. It was connected to a statistical model in Helland (1990). Briefly, this model can be formulated as follows: Let $\Sigma_x$ be the covariant matrix of the $p$ explanatory variables $x_i; i=1,...,p$, assumed to be random, and let $\{\bm{d}_i\}$ be the eigenvectors of $\Sigma_x$. Decompose the regression vector $\bm{\beta}$ of the predictor variable $y$ upon $\bm{x}=(x_1,...,x_p)$ as
\begin{equation}
\bm{\beta}=\sum_{i=1}^p \gamma_i \bm{d}_i,
\label{regrbeta}
\end{equation}
and then introduce the following hypothesis for some $m<p$:

\[H_m:\ \mathrm{There\ are\ exactly\ } m\ \mathrm{nonzero\ terms\ in\ (\ref{regrbeta}).}\]

There are two mechanism by which the number of terms can be reduced: 1) Some terms are really zero; 2) There are coinciding eigenvalues of $\Sigma_x$, and then the eigenvectors may be rotated in such a way that there is only one in the relevant eigenspace that is along $\beta$.

Then the following is proved in Helland (1990): The parametric version of the partial least squares regression algorithm stops after $m$ steps under the hypothesis $H_m$. Using the resulting partial least squares regression for prediction seems to give a good solution to the collinearity problem.

In Helland (2024f) this is studied further. Among other results, one can prove the following: (Theorem 5 in op. cit.) Using a least squares criterion, the partial least squares model under certain technical conditions gives the best possible model reduction for linear prediction, using an expected least squares criterion.

In discussing this and related results, it turns out to be of some relevance to use results from the foundation of quantum theory, in particular Theorem 1 from Subsection 2.2. Specifically, let $\phi=(\beta, \Sigma_x, \sigma^2 )$ be the full parameter of the model, where $\sigma^2 =\mathrm{Var}(y|x_i; i=1,...,p)$, let $\theta=\theta(\phi)$ be the model reduced $\beta$ under the hypothesis $H_m$, and let $\eta=\eta(\phi)$ be any other $m$-parametric model reduction of $\beta$. Then it is shown in Helland (2024f) that the assumptions of Theorem 1 are satisfied.

Using this, it is shown: The technical condition ensuring that the PLS regression model ($\theta$) is better than the arbitrary reduction $\eta$ can be replaced by a condition involving a statistician $B$ that has an non-informative prior on $\eta$.

Furthermore, assume Postulate 5 and Postulate 6. Then the general version of the Born formula holds. In Postulate 6 we may specify the superior actor $D$ to represent general scientific ideals connected to any statistician making the statistical analysis. 

\subsection{Continuous parameters and complementarity}

The discussion in this subsection must be considered tentative. There are mathematical issues that should be resolved. Some of these issues are cleared up in Helland (2026a).

In the traditional approach to quantum mechanics, the Hilbert space is directly determined by the variable considered. If this is the position of a particle, say, the Hilbert space is $L^2 (\mathbb{R}, d\mu)$, where $\mu$ is Lebesgue measure, and the operator corresponding to position $x$ is just a multiplication by $x$.

Similarly, for a continuous statistical parameter $\theta$ that varies over the whole space, and where the relevant group is the translation group, we can take the Hilbert space to be $L^2 (\mathbb{R}, d\mu)$, and take the operator $A^\theta$ corresponding to $\theta$ to be multiplication of $f\in L^2 (\mathbb{R}, d\mu)$ by $\theta$.

This also determines the operator for any $\xi =\xi(\theta)$: By the spectral theorem we have $A^\theta =\int_{\sigma^{A^\theta}}\lambda dE^{A^\theta}$, which gives $A^\xi =\int_{\sigma^{A^\theta}}\xi(\lambda) dE^{A^\theta}$. 

Complementarity is a notion due to Niels Bohr, who called the position and momentum of a particle complementary variables. In my theory in a statistical context two parameters are called complementary if they are really different and both are maximal accessible variables. By Theorem 1, the existence of two such complementary parameters is the essential basis for the development of quantum phenomena in a statistical setting.

Subsection 3.1 gave a setting where two discrete complementary parameters implied quantum probability as a possible prior. For continuous parameters, the theory is more complicated. The point is that the Hilbert space $L^2 (\mathbb{R}, d\mu)$ is not separable and does not have a countable basis. On many occasions, also in this article, it may nevertheless be useful to think in terms of a finite set of basis vectors. This corresponds to parameters $\theta$ and $\eta$ taking a finite number of values. Continuous parameters may be approximated by such finite-valued parameters. For mathematicians, a direct strictly precise theory is given in Hall (2013).
\bigskip

\underline{Example}

Consider a modified version of the example from Subsection 3.1. Assume that in the rash-medicine illustration, we really are able to measure the difference in improvement on the sides of each patient's back, and thus in one set of measurements get an estimate of $\zeta^a = \mu_a - \frac{1}{3} (\mu_b +\mu_c + \mu_d)$ and in another set of measurements get an estimate of $\zeta^b = \mu_b - \frac{1}{3} (\mu_a +\mu_c + \mu_d)$. These are contrasts, but not orthogonal contrasts.

We are interested in the contrast $\zeta^b$, but relative to a particular population for which we have information about the contrast $\zeta^a$. This information is obtained from a previous experiment that has been analyzed by either frequentist or Bayesian methods. In the first case we have obtained a confidence distribution of the contrast $\zeta^a$, and in the last case a posterior probability distribution. In either case, we possess now a probability density $p^a (\zeta)$ for $\zeta^a$. 
We will use this to find a prior for the experiment on $\zeta^b$, relevant for the resulting population.

Now we use Theorem 1. The assumptions may be shown to be satisfied, see below, with $\theta=\zeta^b$, $\eta=\zeta^a$ and $G$ equal to the translation group on $\theta$, and $K$ equal to the translation group on $\phi=(\mu_a, \mu_b, \mu_c, \mu_d)$. The result is two symmetric operators, $A^a$ corresponding to $\zeta^a$ and  $A^b$ corresponding to $\zeta^b$. We can take $A^a$ to be of the simple multiplication form. The operators will be self-adjoint. Using the spectral theorem for $A^a$ ($A^a = \int \zeta dE^a (\zeta)$) and the probability density $p^a (\zeta)$, we find a density operator $\rho^a =\int p(\zeta)dE^a(\zeta)$, and the spectral theorem for $A^b$ ($A^b = \int \eta dE^b (\eta)$) gives a projection operator $\Pi_B = \int_B dE^b (\eta)$ for each Borel set $B$. Then from Born's formula a prior for $\zeta^b$ is given by 
\begin{equation}
P(\zeta^b \in B |\rho^a) = \mathrm{trace}(\rho^a \Pi_B).
\label{contrastB}
\end{equation}

It is crucial for this example that both the parameters $\zeta^a$ and $\zeta^b$ can be seen to be maximal as accessible parameters; see Definition 1 of Subsection 2.2. They belong to different experiments and must be maximal in these experiments: For instance, any parameter $\lambda$ for the first experiment such that $\zeta^a$ is a function of $\lambda$ which is not bijective must be inaccessible, not possible to estimate from the available data. This is a rather strong requirement.

One such potential $\lambda$ will be the vector $(\mu_a, \frac{1}{3}(\mu_b+\mu_c+\mu_d))$, another will be the vector $(\mu_a, \mu_b, \mu_c, \mu_d)$. These must be inaccessible; it is only possible to measure contrasts/ differences between sides of the back. (This can be seen as a somewhat strange requirement, but it is here crucial for my arguments.) Furthermore, one must argue that priors should only depend upon accessible parameters. Then, a quantum prior determined by (\ref{contrastB}) can be motivated for the second experiment.

Note that in this example, if we take the basic inaccessible variable as  $\phi=(\mu_a, \mu_b, \mu_c, \mu_d)$, and $K$ as the translation group acting on $\phi$, then the contrast function $\theta=\zeta^b = \mu_b - \frac{1}{3} (\mu_a +\mu_c + \mu_d)$ is not permissible with respect to $K$. (See Definition 2 of Subsection 3.1.) The group $G$ acting on $\theta$ is not induced by the full group $K$.

In the discussion above, I had assumed that we really had knowledge about $\zeta^a$ for each unit, so that this was available for the selection of units. For the more realistic case, see below.
\bigskip

The essence of this example can be generalized. Assume a statistical model $P(z|\phi)$ with density $p(z|\phi)$ with some large parameter space $\Omega_\phi$. Let $K$ be a group acting upon $\Omega_\phi$, and let $\theta=\theta(\phi)$ be some focus parameter. Assume that $\theta$ is a maximal accessible parameter in the given situation, that is, 1) It can be estimated using some estimation principle; 2) If $\theta=f(\xi)$ for a non-invertible function $f$, then the parameter $\xi$ cannot be estimated. Assume that there is a transitive group $G$ with a trivial isotropy group acting on $\Omega_\theta$. Consider some experimental units, and make an experiment in accordance with the model on these units. This gives a Bayesian posterior or a confidence distribution (more generally, a fiducial distribution; see  Murph et al., 2024) with density $p(\theta)$. In the Bayesian case, it is natural, if possible, to use as a prior the invariant measure associated with the group $G$.

Then do a new experiment on a selected set of units, selected according to the probability distribution $p(\theta)$. Let $\eta=\eta(\phi)$ be another maximal accessible parameter, a focus parameter on the second experiment and essentially different from $\theta$. Then, according to Theorem 1, there exists a Hilbert space $\mathcal{H}$, and two symmetric operators $A^\theta$ and $A^\eta$ in $\mathcal{H}$, one associated with $\theta$ and one with $\eta$. Let $A^\theta= \int \theta dE^\theta(\theta)$ be the spectral decomposition of $A^\theta$, and define $\rho^\theta=\int p(\theta)dE^\theta (\theta)$. Then it can be argued that a prior $\pi$ for the second experiment should be chosen such that $\pi(B)=\mathrm{trace}(\rho^\theta \Pi_B)$, where $\Pi_B$ is the projection operator defined by $\Pi_B = \int_B dE^\eta (\eta)$, with $\{E^\eta\}$ chosen such that $A^\eta = \int \eta dE^\eta (\eta)$.

  All this assumes that units can be chosen by values of $\theta$ that are really known. If not, we define $u=\widehat{\theta}$, where $\widehat{\theta}$ is the chosen estimator of $\theta$, and we must replace $\rho^\theta$ by $\rho^u =\int r(u) dE^u (u)$,  where $r(u)=\int q(u|\theta)p(\theta)d\theta$ with $q$ being the density in the distribution of the estimator, assumed to only depend upon $\theta$, and where $\{E^u \}$ is found from the spectral distribution of $A^u =\int u q(u|\theta)dE^\theta(\theta)$. Units are then chosen from data $z$ of the first experiment according to the density $r(u(z))$.

Similar discussions can in principle be made in very complicated statistical models. For many such models, the groups $G$ and $K$ of Theorem 1 can be defined. An example from the design of experiments where such groups can be defined is the model for randomized experiments discussed by Bailey (1991). But to discuss examples where links to quantum theory can be found, one has to have a basic inaccessible parameter $\phi$, and two really different maximal accessible (complementary) parameters.

An even more general class of statistical models is discussed by McCullagh (2002) using category theory. Group theory can be seen as a special case of category theory, and using this, examples with groups $G$ and $K$ can be found. But again, it is a challenge to find applied examples with two different maximal accessible parameters.

It is obvious that more research in this area is required.

\subsection{Machine Learning}

The literature on Artificial Intelligence, in particular Machine Learning, has exploded in recent years. For the purpose of this article, I will focus on a simple neural network with one hidden layer and a single output, as described from a statistical point of view in Efron and Hastie (2021). Here, assume $p$ predictors (features) $\bm{x}=(x_1, x_2,...,x_p )$ which for simplicity are centered on zero expectation, $k$ hidden units $a_l = g(\sum_{j=1}^p w_{jl}x_j)$ ($l=1,...,k)$ and output $z=h(\sum_{l=1}^k v_l a_l)$, where $g$ and $h$ are non-linear, monotonically increasing functions satisfying $g(0)=0$ and $h(0)=0$, and $\{w_{jl}\}$ and $\{v_l\}$ are the parameters of the model, the weights. The $\{\ x_j \}$ and $y$ are observed on $n$ units, and our task - the learning of the network - is to estimate the weights. In many applications, $n$ is very large, and procedures such as backpropagation are used. I will here also consider the case where $n$ is moderate, perhaps smaller than $p$. Then a model reduction may be called for.

Before discussing this, I  take a brief look at some of the recent literature concerning links between machine learning and quantum theory.

  In their abstract, Dunjko and Briegel (2019) mention 3 points: 1) Quantum computing is finding vital applications in providing speed-ups for machine learning problems. 2) Machine learning may become instrumental in advanced quantum technologies. 3) One can consider quantum generalizations of learning and artificial intelligence concepts. 
  
Op. cit. is a review article with many references to recent papers. Quantum models of relevance to machine learning are discussed in detail. Historically, the first such model was the quantum Turing machine (Deutch, 1985), but there are many more modern models. In general, machine learning can be divided into supervised and unsupervised learning. In supervised learning we start with a training set $(y_i, \bm{x}_i), i=1,...,n$, where $\bm{x}_i$ is a vector, and the task is to predict $y$ from $\bm{x}$ on a new unit, so multiple regression can be seen as a special case of machine learning. Quantum information is a more general concept, where data are replaced by quantum states. This is a large area with independent literature. Again, specializing to the multiple regression case, one can mention works by Wiebe et al. (2012), Wang (2017), and Schuld et al. (2016).

In two recent articles, Wu et al. (2023) and Zhu et al (2023) develop network models that can simultaneously predict multiple quantum properties and the behavior of an unknown quantum process.

Quantum foundations seek to understand and develop the mathematical and conceptual basis for quantum theory. Bharti et al. (2020) survey representative works at the interface of machine learning and quantum foundations. Special topics considered are entanglement, Bell-type inequalities and contextuality. It is proposed that neural networks can be seen as `hidden' variable models for quantum systems.

Go back to the simple model introduced in the first paragraph of this Section. We are interested in model reduction. The whole neural net is usually learned by gradient descent (Efron and Hastie, 2021, Charniak, 2018). For the purpose of model reduction, I will concentrate on a single perceptron
\begin{equation}
a=g(\sum_{j=1}^p w_j x_j).
\label{perceptron}
\end{equation}

I now as a model, assume that the features $\bm{x} = (x_1,..., x_p)$ have a random distribution with expectation $\bm{0}$ and covariance matrix $\bm{\Sigma}$, assumed to be positive definite. Expand $\bm{w}=(w_1,...,w_p)$ in terms of eigenvectors $\bm{d}_i$ of $\bm{\Sigma}$
\begin{equation}
\bm{w}=\sum_{i=1}^p \gamma_i \bm{d}_i.
\label{wexp}
\end{equation}

Then, completely in analogy with partial least squares regression (Subsection 3.3), introduce the model reduction ($\theta_m$)

\[H_m:\ \mathrm{There\ are\ exactly\ } m\ \mathrm{nonzero\ terms\ in\ (\ref{wexp}).}\]

The theory of Helland (2024f) carries over. We can represent $H_m$ with the parameter $\theta_m =(\gamma_1 \bm{d}_1,...,\gamma_m \bm{d}_m)$, which is a function of $\phi=(\bm{w},\bm{\Sigma})$. Assume any other model reduction $\eta_m$ to $m$ terms. Assume a non-informative probability distribution of $\eta_m$. Then, it can be proved that by a least squares criterion, $\theta_m$ gives a better model reduction  than $\eta_m$.

In proving this in op. cit., essential use was made of a joint quantum model for $\theta_m$ and $\eta_m$. There, and here, the assumptions of Theorem 1 (Subsection 2.2) can be shown to be satisfied. The group $G$ acting on $\theta$ is given by orthogonal transformations of the $\bm{d}_j$'s and $\gamma_j \mapsto \alpha_j\gamma_j$ $(j=1,2,...,m)$. It is convenient to let the group $K$ on $\phi$ be defined by orthogonal transformations of all the $\bm{d}_i$'s $(i=1,...,p)$ and by $\gamma_i \mapsto g(\alpha_i \gamma_i)$. Then the orbits of the group $K$ are given by $m$ and the hypothesis $H_m$. (Theorem 2 in Helland (2024f)). This is the reason why I have chosen the constraint $g(0)=0$.

To carry out the model reduction for the perceptron in theory, and also in practice, the whole literature on partial least squares can be taken over if we base ourselves on $\bm{x}$ and $y=g^{-1}(a) [=\sum_{j=1}^p w_j x_j]$. The theoretical population algorithm is given in Appendix 1 of Helland (2024f). In practice, we have data on $n$ copies of $(\bm{x},y)$, and in the algorithm, theoretical variances and covariances must be replaced by estimated variances and covariances. The size $m$ can be determined by cross-validation.

Then all this should be incorporated into the algorithm for the whole neural network. Look again at the simple model defined at the beginning of this Subsection. We are given data $(x,z)$ on $n$ units, where all these variables are centered by their means. We use a feedforward procedure, which means that we start the estimation by first looking at the transition from the $x$-data to the variables $a_l$, then from $a_l$ to $z$. The now well-known procedures here can be found in the machine learning literature. Each of these steps must now be replaced by a series of steps of the type described in the previous paragraph. I omit the details here, since the purpose of the present article is to introduce ideas.

 A model reduction of this kind can be expected to give an advantage when there is not too much data ($n$) compared to the number $p$ of variables, or perhaps more accurately, as shown in the partial least squares case by recent asymptotics by Cook and Forzani (2019), in the abundant case where many predictors $x$ contribute information about the response $z$, often correlated information.

\section{Concluding remarks}

The social scientist Ralph D. Stacey once said: `Culture is a set of attitudes, opinions and convictions that a group of humans share, about how one shall behave against each other, how things shall be evaluated and done, which questions that are important and which answers that are accepted. The most important elements of culture are unconscious, and cannot be forced upon us from the outside.'

From this perspective, statistical theory and quantum theory, as they have functioned up to now, may be seen as connected to separate cultures. It is hoped that this article may help to bridge the gap between these two cultures.

The investigations here started with Helland (2022c), where mathematical models in various sciences were discussed from several points of view. With the present article, this discussion can be said to lead to concrete results.

Of course, there are differences between models in quantum theory on the one side and statistically related models on the other side. It is important that quantum models are always seen in a context, and that they often are related to an observer or a group of communicating observers. Contextual quantum measurements have been discussed from several points of view by Khrennikov (2024). By contrast, statistical models are more universal. But this does not contradict the fact that quantum-like models are ubiquitous, cf. Khrennikov (2010, 2023).

\section*{Acknowledgements}

I really want to thank Andrei Khrennikov for his yearly, very enlightening conferences on quantum foundations and related   
topics. I have learned a lot by attending a few of these conferences. I also want to thank Christopher Fuchs and Richard Gill for their patience in trying to understand my basic message. Finally I am grateful to Wolfgang Tiefenbrunner for making me aware of the works by Hans De Raedt et al. I am grateful to Solve S\ae b\o\  and Trygve Alm\o y for discussions, and I am grateful to Gudmund Hermansen for doing numerical calculations in connection to the experiment in Subsection 3.1.

\section*{Declarations}

\subsection*{Ethical approvement and consent to participate}

Not applicable.

\subsection*{Consent for publication}

Consent for publication of this article is given.

\subsection*{Funding}

There is no funding behind this article.

\section*{References}

$\ \ \ \ $Bailey, R.A. (1991). Strata for randomized experiments. \textit{J, Roy. Statist. Soc, B} 51(1), 27-68.
\smallskip

Barndorff-Nielsen, O.E., Gill, R.D., and Jupp, P.E. (2003). On quantum statistical inference. \textit{J. R. Statist. Soc. B} \textbf{65} (4), 1-31.
\smallskip

Berger, J.~O. and Wolpert, R.~L. (1988). \textit{The likelihood principle.} Institute of Mathematical Statistics, Hayward, CA.
\smallskip

Bharti, K., Haug, T., Vedral, V., and Kwek, L.-C. (2020). Machine learning meets quantum foundations: A brief survey. \textit{AVS Quantum Sci.} 2, 034101.
\smallskip

Busch, P. (2003). Quantum states and generalized observables: A simple proof of Gleason's Theorem. \textit{ Physical Review Letters} 91(12), 120403.
\smallskip

Busemeyer, J.R. and Bruza, P.D. (2012). \textit{Quantum Models of Cognition and Decision.} Cambridge University Press, Cambridge.
\smallskip

Caves, C. ~M., Fuchs, C.~A., and Schack, R. (2002). Quantum probabilities as Bayesian probabilities. \newblock \textit{Physical Review, A65}, 022305.
\smallskip

Charniak, E. (2018). \textit{Introduction to Deep Learning.} The MIT Press, Cambridge, Massachusetts.
\smallskip

Cook, R.D. \& Forzani, L. (2019). Partial least squares prediction in high-dimensional regression. \textit{Ann. Stat.} 47 (2), 884-908.
\smallskip

De Raedt, H., Katsnelson, M.J., and MIchelson, K. (2014). Quantum theory as the most robust description of reproducible experiments. \textit{Annals of Physics} 347, 45-73.
\smallskip

De Raedt, H., Katsnelson, M.J., and MIchelson, K. (2016). Quantum theory as plausible reasoning applied to data obtained by robust experiments. \textit{Phil. Trans. Roy. Soc. A} \textbf{374}, 20150233.
\smallskip

De Raedt, H., Katsnelson, M.J., Jattana, M.S., Mehta, V., Willsch, M., Willsch, D., MIchelson, K. and Jin, F. (2023). Einstein-Podolsky-Rosen-Bohm experiments: A discrete data driven approach. \textit{Annals of Physics} \textbf{453}, 169314.
\smallskip

Deutch, D. (1985). Quantum theory, the Church-Turing principle and the universal quantum computer. \textit{Proc. R. Soc. A} 400, 97-117.
\smallskip

Dunjko, V. and Briegel, H.J. (2019).
Machine learning $\&$ artificial intelligence in the quantum domain: a review of recent progress. \textit{Rep. Prog. Phys.} \textbf{81}, 074001.
\smallskip

Efron, B. and Hastie, T. (2021). \textit{Computer Age Statistical Inference} 2. Edition. Cambridge University Press, Cambridge
\smallskip

Frieden, B.R. (2004). \textit{Science from Fisher Information: A Unification.} Cambridge University Press, Cambridge.
\smallskip

Fuchs, C.A. (2003). Quantum mechanics as quantum information, mostly. \textit{J. Mod. Opt.} \textbf{50}, 987-1023.
\smallskip

Fuchs, C.A. (2010). QBism, the perimeter of quantum Bayesianism. arXiv: 1003.5209 [quant-ph].
\smallskip

Hall, B.C. (2013).  \textit{Quantum Theory for Mathematicians.}  Graduate Texts in Mathematics, \textbf{267}, Springer, Berlin.
\smallskip

Haven, E. and Khrennikov, A. (2013). \textit{Quantum Social Science.} Cambridge University Press, Cambridge.
\smallskip

Haven, E. and Khrennikov, A. (2016). Statistical and subjective interpretations of probability in quantum-like models in cognition and decision-making. \textit{J. Mathematical Psychology} 74, 82-91.
\smallskip

Helland, I.S. (1990). Partial least squares regression and statistical models. \textit{Scand. J. Statist} 17, 97-114.
\smallskip

Helland, I.S. (2001). Reduction of regression models under symmetry. \textit{Contemp. Math.} 287. 139-154.
\smallskip

Helland, I.S. (2004). Statistical inference under symmetry. \textit{Int. Stat. Rev.} 72 (3), 409-422.
\smallskip

Helland, I.S. (2006). Extended statistical modeling under symmetry: The link toward quantum mechanics. \textit{Ann. Statist.} 34 (1), 42-77.
\smallskip

Helland, I.S, (2021).
\textit{Epistemic Processes. A Basis for Statistics and Quantum Theory.} 2. edition. Springer, Berlin.
\smallskip

Helland, I.S, (2022a). On reconstructing parts of quantum theory from two related maximal conceptual variables.  \textit{International Journal of Theoretical Physics} 61, 69.
\smallskip

Helland, I.S. (2022b). The Bell experiment and the limitation of actors. \textit{Foundations of Physics} 52, 55
\smallskip

Helland, I.S. (2022c). On the diversity and similarity of mathematical models in science. \textit{American Review of Mathematics and Statistics} 10 (1), 1-10.
\smallskip

Helland, I.S. (2023a). A new foundation of quantum decision theory. arXiv:2310.12762 [quant-ph]
\smallskip

Helland, I.S. (2023b). Possible connections between relativity theory and a version of quantum theory based upon theoretical variables. arXiv: 2305.15435 [physics.hist-ph]
\smallskip

Helland, I.S. (2023c). A simple quantum model linked to decisions. \textit{Foundations of Physics} 53, 12.
\smallskip

Helland, I.S. (2023d). On the Bell experiment and quantum foundation. arXiv: 2305.05299 [quant-ph]. \textit{J. Mod. Appl. Phys.} 6 (2), 1-5.
\smallskip

Helland, I.S, (2024a).
An alternative foundation of quantum theory. \textit{Foundations of Physics} \textbf{54}, 3, arXiv: 2305.06727 [quant-ph].
\smallskip

Helland, I.S. (2024b). On probabilities in quantum mechanics. arXiv: 2401.17717 [quant-ph]. \textit{APL Quantum}
1, 036116. https://doi.org/10.1063/5.0218982.
\smallskip

Helland, I.S. (2024c). A new approach toward the quantum foundation and some consequences. \textit{Academia Quantum} 1. https://doi.org/AcadQuant7282.
\smallskip

Helland, I.S. (2024d). Some mathematical issues regarding a new approach towards quantum foundation. arXiv: 2411.13113 [quant-ph].
\smallskip

Helland, I.S. (2024e). On the foundation of quantum decision theory. arXiv: 2310.12762 [quant-ph]. 
\smallskip

Helland, I.S. (2024f) On optimal linear prediction. arXiv: 2444412.19186 [math.ST]
\smallskip

Helland, I.S. (2026a) The final version of a recent approach towards quantum foundation. arXiv: 2604.05675 [quant-ph]
\smallskip

Helland, I.S. (2026b) On the Born rule in a new quantum approach. arXiv: 2609.24269 [quant-ph]
\smallskip

Helland, I.S. and Parthasarathy, H. (2024). \textit{Theoretical Variables, Quantum Theory, Relativistic Quantum Field Theory, and Quantum Gravity.} Manakin Press, New Dehli.
\smallskip

Jaynes, E.T. (2003). \textit{Probability Theory: The Logic of Science.} Ed.: G.L. Bretthorst. Cambridge University Press, Cambridge.
\smallskip

 Khrennikov, A. (2010). \textit{Ubiquitous Quantum Structure. From Psychology to Finance.} Springer, Berlin.
\smallskip

 Khrennikov, A. (2023). \textit{Open Quantum Systems in Biology, Cognitive and Social Sciences.} Springer Nature Switzerland.
\smallskip

Khrennikov, A. (2024). Contextual measurement model and quantum theory. \textit{Royal Society Open Science} 11, 231953.
\smallskip

Kolmogorov, A.N. (1933). \textit{Grundbegriffe der Wahrscheinlichkeitsrechnung.} Springer, Berlin. English translation: Kolmogorov, A.N. (1956). \textit{Foundations of the Theory of Probability.} 2. edition. Chelsea, New York.
\smallskip

Liu, H. (2024). A quantum Bayes' rule and related inference. \textit{Quantum Information Processing} 23:271, 1-26..
\smallskip

Maccone, L. (2024). Schr\"{o}dinger cats and quantum complementarity. \textit{Foundations of Physics} 54, 17.
\smallskip

McCullagh, P. (2002). What is a statistical model? \textit{Ann. Statist.} 30 (5), 1225-131.
\smallskip

Murph, A.C., Hannig, J. and Williams, J.P. (2024). Introduction to generalized fiducial inference. In: Berger, J., Meng, X.-L., Reid, N.. \& Xie, M. M.-g. (2024). \textit{Handbook of Bayesian, Fiducial, and Frequentist Inference.} Chapman and Hall.
\smallskip

Plotnitsky, A. (2013). \textit{Niels Bohr and Complementarity. An Introduction.} Springer, New York.
\smallskip

Schweder, T. and Hjort, N.L. (2016). \textit{Confidence, Likelihood, Probability. Statistical Inference with Confidence Distributions.} Cambridge University Press, Cambridge.
\smallskip

Schuld, M., Sinaiskyi, I., and Petruccione F, (2016). Prediction by linear regression on a quantum computer. \textit{Phys. Rev. A} 94, 022342.
\smallskip

Steiner, M. and Rendell, R. (2024). Complementary relationships between entanglement and measurement. \textit{Academia Quantum} 1.
\smallskip

Wang, G. (2017). New quantum algorithm for linear regression. \textit{Phys. Rev. A} 96, 012335.
\smallskip

Wiebe, N., Braun, D., and Lloyd, S. (2012). Quantum algorithm for data fitting. \textit{Phys. Rev. Lett.} 109, 050505.
\smallskip

Wu, Y.-D., Zhu, Y., Wang, Y., and Chirabella, G. (2023). Learning and discovering quantum properties with multi-task neural networks. arXiv: 2310.11807 [quant-ph].
\smallskip

Zhu, Y., Wu, Y.-D., Liu, Q. Wang, Y., and Chirabella,G. (2023). Predictive modelling of quantum process with neural networks. arXiv: 2308.08815 [quant-ph].
\smallskip

\end{document}